\documentclass[numberedappendix]{emulateapj}
\pdfoutput=1

\usepackage{epsfig, natbib}
\tightenlines

\countdef\decade=200
\advance\decade by \year
\countdef\hours=201	
\advance\hours by \time
\divide\hours by 60
\countdef\mins=202
\advance\mins by \hours
\multiply\mins by 60
\multiply\hours by 100
\countdef\miltime=203
\advance\miltime by \hours
\advance\miltime by \time
\advance\miltime by -\mins

\def\mathbi#1{\textbf{\em #1}}

\newcommand{\msun}{M_{\odot}}

\newcommand{\tff}{t_{\rm ff}}
\newcommand{\vecv}{\mathbi{v}}
\newcommand{\vecx}{\mathbi{x}}
\newcommand{\vecp}{\mathbi{p}}
\newcommand{\krho}{k_{\rho}}

\defcitealias{krumholz07a}{KKM07}
\defcitealias{krumholz08a}{KM08}

\begin{document}

\title{Radiation Feedback, Fragmentation, and the Environmental Dependence of the Initial Mass Function}

\slugcomment{Accepted to the Astrophysical Journal}

\shorttitle{Radiation Feedback and the IMF}
\shortauthors{Krumholz et al.}

\author{
        Mark R. Krumholz\altaffilmark{1},
        Andrew J. Cunningham\altaffilmark{2},
        Richard I. Klein\altaffilmark{2, 3}, and
        Christopher F. McKee\altaffilmark{3,4}}

\altaffiltext{1}{Department of Astronomy and Astrophysics,
         University of California, Santa Cruz, CA 95064;
         krumholz@ucolick.org}
\altaffiltext{2}{Lawrence Livermore National Laboratory, P.O. Box 808, L-23, Livermore, CA 94550}
\altaffiltext{3}{Department of Astronomy and Astrophysics, University of California, Berkeley,
Berkeley, CA 94720}
\altaffiltext{4}{Department of Physics, University of California, Berkeley,
Berkeley, CA 94720}




\begin{abstract}
The fragmentation of star-forming interstellar clouds, and the resulting stellar initial mass function (IMF), is strongly affected
by the temperature structure of the collapsing gas. Since radiation feedback from embedded stars can modify this as collapse proceeds, feedback plays an important role in determining the IMF. However, the effects and importance of radiative heating are likely to depend strongly on the surface density of the collapsing clouds, which determines both their effectiveness at trapping radiation and the accretion luminosities of the stars forming within them. In this paper we report a suite of adaptive mesh refinement radiation-hydrodynamic simulations using the ORION code in which we isolate the effect of column density on fragmentation by following the collapse of clouds of varying column density while holding the mass, initial density and velocity structure, and initial virial ratio fixed. We find that radiation does not significantly modify the overall star formation rate or efficiency, but that it suppresses fragmentation more and more as cloud surface densities increase from those typical of low mass star-forming regions like Taurus, through the typical surface density of massive star-forming clouds in the Galaxy, up to conditions found only in super star clusters. In regions of low surface density, fragmentation during collapse leads to the formation of small clusters rather than individual massive star systems, greatly reducing the fraction of the stellar population with masses $\ga 10$ $\msun$. Our simulations have important implications for the formation of massive stars and the universality of the IMF.
\end{abstract}


\section{Introduction}
\label{intro}

The stellar initial mass function (IMF) is determined by the fragmentation of gas clouds into progressively smaller pieces as they collapse, and the characteristic fragment mass scale is thought to be proportional to the Jeans mass. However, in an infinite, isothermal gas cloud the Jeans analysis does not pick out a unique mass scale for fragmentation. Since the Jeans mass varies as $M_J \propto \rho^{-1/2} T^{3/2}$, where $\rho$ and $T$ are the gas density and temperature, in an isothermal cloud $M_J$ reaches arbitrarily small values as the cloud collapses and $\rho$ increases, allowing fragmentation to proceed to arbitrarily small scales. In terms of numerical simulations, this lack of a natural fragmentation scale in isothermal clouds is reflected in the fact that the amount of fragmentation is ultimately resolution-dependent \citep[e.g.][]{martel06}, and that an isothermal simulation in a periodic box can always be rescaled so as to give the fragments that form an arbitrary mass, while maintaining a fixed virial parameter, Mach number, and number of initial Jeans masses \citep[e.g.][]{offner08b}.

For this reason, any explanation of the fragmentation scale of molecular clouds and thus the stellar IMF requires a deviation from uniform, isothermal flow. However, the nature of this deviation remains controversial. In turbulent fragmentation models \citep{padoan02, hennebelle08b, hennebelle09a}, the shape of the IMF is determined by the properties of turbulence, but the overall mass scale is set by computing the Jeans mass at a density proportional to the mean density\footnote{In models of this sort the relevant density is usually taken to be the density times the square of the Mach number, which \citet{krumholz05c} point out is equal to the Jeans mass computed at a pressure equal to the mean ram pressure.} either in an entire giant molecular cloud or in some smaller, fiducial star-forming gas clump. While there is significant support from numerical simulations that a process of this sort operates \citep{padoan07a}, and observations indicate that the Jeans scale is imprinted in CO clumps \citep{blitz97a}, this model has two significant gaps. First, since molecular clouds have complex structures that span a large range of densities, it is not obvious how to define the star-forming region over which the mean density should be computed. Second, once a gravitationally-bound collapsing object is formed in these models, it is unclear why it should not fragment even further as it collapses, since the bound object now defines a new cloud with a higher mean density, and thus a smaller fragmentation scale. Indeed, purely hydrodynamic simulations of isolated massive cores find exactly this behavior \citep{dobbs05}.

In contrast, in non-isothermal fragmentation models the characteristic mass scale is introduced via a small deviation from isothermality that occurs at some density, which then sets the density and temperature that enter into the Jeans mass. The necessary kink in the equation of state may arise in several ways. It can come from from a transition between optically thin and optically thick conditions as gas collapses \citep[e.g.][]{masunaga98, masunaga00, bonnell06d}; the opacity limit for fragmentation, first proposed by \citet{low76a}, is one example of such a transition, albeit at a mass scale of $0.004$ $\msun$ \citep{whitworth07a}, too low to be relevant for the bulk of stars. A second possible origin for non-isothermality is the density-dependent interaction between molecular cooling, cosmic ray heating, and dust-gas coupling \citep{larson05, elmegreen08a}. That non-isothermality of this sort can affect fragmentation also has support from numerical simulations \citep{jappsen05}, but these models too face difficulties. Their procedure for computing the characteristic density and temperature relies on an effective equation of state based on average rates of radiative heating and cooling. This ignores the large spatial and temporal variations in the radiation field in star-forming clouds.

These difficulties have led to a renewed focus on the potential importance of another mechanism for setting a characteristic mass scale: radiation feedback. Young stars radiate prodigiously, due to accretion at low masses and via Kelvin-Helmholtz contraction and nuclear burning at higher masses, and these effects can heat the gas around them, raising the Jeans mass. Analytically, \citet{krumholz06b} and \citet[hereafter KM08]{krumholz08a} show that stellar feedback should strongly suppress fragmentation by raising the temperature in star-forming clouds. Numerical simulations of high-mass star (\citealt{krumholz07a}, hereafter KKM07) and low-mass \citep{offner09a} star formation confirm this conclusion. Under some circumstances feedback effects completely swamp the subtle changes in the equation of state on which the non-isothermal fragmentation models rely.

While the importance of radiation feedback has been recognized, its relative importance in different star-forming environments is only starting to be considered. Analytic models by \citetalias{krumholz08a} suggest that effectiveness of feedback will depend on the column density of the star-forming cloud, which determines its ability to trap protostellar radiation. The effect is strong only when $\Sigma \ga 1$ g cm$^{-2}$. Observed star-forming clouds have surface densities ranging from $\Sigma \sim 0.1$ g cm$^{-2}$ in diffuse clouds such as Perseus and Ophiuchus \citep{evans09a} to $\Sigma \sim 1$ g cm$^{-2}$ in typical Galactic regions of massive star formation \citep{shirley03a, faundez04a, fontani05a} to $\Sigma \sim 10$ g cm$^{-2}$ or more in extragalactic super star clusters \citep{turner00a, mccrady07a}, 
and \citet{mckee02, mckee03a} first pointed out that massive stars seems to form only at the high surface density end of this distribution. However, numerical simulations thus far have not explored this parameter space systematically. \citetalias{krumholz07a} find that feedback suppresses the fragmentation of massive protostellar cores with $\Sigma \sim 1$ g cm$^{-2}$, while \citet{offner09a} find a similar but significantly weaker suppression of fragmentation at $\Sigma\sim 0.1$ g cm$^{-2}$.

The only comparisons of fragmentation with radiation in clouds of varying surface density performed thus far are those of \citet{bate09a} and \citet{urban10a}. Both of these authors consider surface densities $\la 0.4$ g cm$^{-2}$, well below \citetalias{krumholz08a}'s analytically-predicted threshold for fragmentation, and well below the observed column density in the typical region of star cluster formation in the Galaxy. Furthermore, neither include radiative transfer in a way that is appropriate to study suppression of fragmentation in dense regions. In Bate's simulations, gas can radiate only up to the point where it is captured by a sink particle. The radiation it releases when it accretes onto the stellar surface is neglected. \citet{offner09a} find that this leads \citeauthor{bate09a} to underestimate the energy budget available for heating the gas by a factor of 20. Urban et al.'s simulations include radiation from stars but not radiation produced by either compression or viscous dissipation, although \citeauthor{offner09a}'s results suggest that this approximation is not bad once stars begin heating the gas. However, Urban et al.\ also rely on a pre-computed spherically-symmetric profile that does not account for deviations from spherical symmetry in the surrounding density field. This is reasonable for the low optical depth clouds and relatively large length scales they consider, but it is questionable for dense cores and on small spatial scales, where non-symmetric shielding effects can be important -- for example accretion disks are typically colder than the gas above or below them, due to their large optical depths, and Urban et al.'s approach would not capture this effect.

The goal of this paper is to fill that gap by studying molecular cloud fragmentation and the stellar mass function in different star-forming environments while self-consistently taking into account the effects of stellar radiation feedback. We perform a controlled experiment by running a series of simulations of collapsing, turbulent gas clouds in which we hold fixed the initial cloud mass, temperature, virial ratio, and turbulent velocity field, while varying the cloud surface density. To ensure that the results are robust, we are careful to hold numerical aspects of the calculations fixed as well. Each simulation uses the same numerical method, the same criteria for refinement, and the same maximum resolution, so that any differences in outcome should be solely the result of radiative effects. In Section \ref{method} we describe the numerical method and initial conditions we use for these simulations. In Section \ref{results} we report the results, and study how fragmentation varies with initial surface density. Finally, in Section \ref{discussion} we discuss the implications of our results for massive star formation and for the IMF more generally, and we summarize in Section \ref{conclusion}.

\section{Numerical Method and Initial Conditions}
\label{method}

\subsection{Equations and Solution Algorithms}

Our numerical method is identical to that of \citet{krumholz09c}, and we give an extensive description in the Supporting Online Material of that paper, so we only summarize our methods briefly here. Our simulations use the parallel adaptive mesh refinement (AMR) radiation-hydrodynamics code ORION. In ORION, the gas plus radiation fluid is represented at every grid point by a vector of conserved quantities $(\rho, \rho \vecv, \rho e, E)$, where $\rho$ is the density, $\vecv$ is the velocity, $e$ is the specific non-gravitational energy (including kinetic and thermal), and $E$ is the radiation energy density. The computational domain also contains an arbitrary number of point mass ``star" particles, each of which is characterized by a position $\vecx$, a mass $M$, a momentum $\vecp$, and a luminosity $L$.
The code updates these quantities by solving the equations of radiation-hydrodynamics plus gravity in the conservative, mixed-frame form \citep{mihalas82}, retaining terms to order $v/c$ accuracy, and using the flux-limited diffusion approximation to represent the radiation flux \citep{krumholz07b}. The equations for the gas are
\begin{eqnarray}
\label{masscons}
\frac{\partial}{\partial t}\rho & = & - \nabla\cdot(\rho\vecv) - \sum_i \dot{M}_i W(\vecx-\vecx_i) \\
\frac{\partial}{\partial t}(\rho \vecv) & = & -\nabla\cdot(\rho \vecv\vecv) - \nabla P - \rho \nabla \phi - \lambda \nabla E
\nonumber \\
& & {} - \sum_i \dot{\vecp}_i W(\vecx-\vecx_i) 
\label{momcons}
\\
\frac{\partial}{\partial t}(\rho e) & = & -\nabla \cdot [(\rho e+P)\vecv] - \rho \vecv \cdot \nabla \phi - \kappa_{\rm 0P} \rho (4 \pi B - c E) 
\nonumber \\
& & {} + \lambda\left(2 \frac{\kappa_{\rm 0P}}{\kappa_{\rm 0R}} - 1\right) \vecv \cdot \nabla E - \sum_i \dot{\mathcal{E}}_i W(\vecx - \vecx_i) 
\label{econsgas}
\\
\frac{\partial}{\partial t}E & = & \nabla \cdot \left(\frac{c\lambda}{\kappa_{\rm 0R} \rho} \nabla E\right) + \kappa_{\rm 0P} \rho (4 \pi B - c E) 
\nonumber \\
& & {} - \lambda \left(2\frac{\kappa_{\rm 0P}}{\kappa_{\rm 0R}} - 1\right) \vecv\cdot \nabla E - \nabla \cdot \left(\frac{3 - R_2}{2} \vecv E\right)
\nonumber \\
& & {}
 + \sum_i L_i W(\vecx - \vecx_i),
\label{econsrad}
\end{eqnarray}
where the summations run over all star particles present, $\dot{M}_i$, $\dot{\vecp}_i$, and $\dot{\mathcal{E}}_i$ are the rates at which mass, momentum, and energy are transferred from gas to star particles, and $W(\vecx-\vecx_i)$ is a weighting function that distributes the transfer over a kernel 4 cells in radius. We calculate these using the \citet{krumholz04} sink particle algorithm, which we summarize below. The corresponding evolution equations for the star particles are
\begin{eqnarray}
\label{starmass}
\frac{d}{dt} M_i &= & \dot{M}_i \\
\label{starpos}
\frac{d}{dt} \vecx_i & = & \frac{\vecp_i}{M_i} \\
\label{starmom}
\frac{d}{dt} \vecp_i & = & -M_i \nabla \phi + \dot{\vecp}_i.
\end{eqnarray}
In these equations, the gravitational potential $\phi$ is given by
\begin{equation}
\label{poisson}
\nabla^2\phi = -4\pi G \left[ \rho + \sum_i M_i \delta(\vecx-\vecx_i)\right].
\end{equation}
The pressure $P$ is given by
\begin{equation}
P = \frac{\rho k_B T_g}{\mu m_{\rm H}} = (\gamma-1) \rho \left(e - \frac{v^2}{2}\right),
\end{equation}
where $T_g$ is the gas temperature, $\mu = 2.33$ is the mean molecular weight for molecular gas of Solar composition, and $\gamma$ is the ratio of specific heats. We adopt $\gamma=5/3$, appropriate for gas too cool for hydrogen to be rotationally excited, but this choice is essentially irrelevant because $T_g$ is set almost purely by radiative effects. The remaining quantities are the comoving frame specific Planck- and Rosseland-mean opacities $\kappa_{\rm 0R}$ and $\kappa_{\rm 0P}$, the Planck function $B = c a_R T_g^4 / (4\pi)$, and the flux limiter $\lambda$ and Eddington factor $R_2$, computed using the \citet{levermore81} approximation:
\begin{eqnarray}
\lambda & = & \frac{1}{R} \left(\mbox{coth} R - \frac{1}{R}\right) \\
R & = & \frac{|\nabla E|}{\kappa_{\rm 0R} \rho E} \\
R_2 & = & \lambda + \lambda^2 R^2.
\end{eqnarray}
We obtain the dust opacities $\kappa_{\rm 0P}$ and $\kappa_{\rm 0R}$ from a piecewise-linear fit to the models of \citet{pollack94}; see \citet{krumholz09c} for the exact functional form. It is worth noting that the Rosseland opacity we use includes absorption but not scattering effects, and as a result is likely something of an underestimate. Using a higher opacity would likely enhance the radiation effect we describe below, as suggested by recent static radiative transfer calculations \citep{dunham10a}.

We solve these equations in four steps. First, the hydrodynamics module updates equations (\ref{masscons}) -- (\ref{econsgas}) using all the terms on the right hand side except those involving star particles or radiation. The update is based on a conservative Godunov scheme with an approximate Riemann solver, and is second-order accurate in time and space \citep{truelove98, klein99}. Second, the gravity module solves the Poisson equation (\ref{poisson}) to update the gravitational potential using a multigrid iteration scheme \citep{truelove98, klein99, fisher02}. Third, the radiation module updates the right-hand sides of equations (\ref{momcons}) -- (\ref{econsrad}) for the terms involving radiation. The module uses the \citet{krumholz07b} operator splitting method, in which the dominant terms describing radiation-gas energy exchange and radiation diffusion are updated using a fully implicit method based on pseudo-transient continuation \citep{shestakov08}, and then the sub-dominant work and advection terms are handled explicitly. The fourth step is the star particle module. This portion of the code updates equations (\ref{masscons}) -- (\ref{starmom}) using the gas-particle exchange terms. The code computes $\dot{M}_i$ by fitting the flow in the vicinity of each particle to a Bondi-Hoyle flow, and then set $\vecp_i$ and $\mathcal{E}_i$ by requiring that, in the frame comoving with the particle, accretion not alter the radial velocity, angular momentum, or temperature of the gas \citet{krumholz04}. The method then updates the luminosity and the internal state of each star based on a simple one-zone protostellar evolution model. Details of the model are given in the Appendices of \citet{offner09a}.

All of these modules operate with an AMR framework \citep{berger84, berger89, bell94} in which we discretize the computational domain onto a series of levels $l=0,1,2,\ldots,L$. The coarsest level is $l=0$, which covers the entire computational domain. Subsequent levels cover sub-regions of the computational domain which are described by the union of rectangular grids. Levels are nested such that any grid on level $l>0$ must be fully contained within one or more grids of level $l-1$. The cell size on level $l$ is $\Delta x_l$, and the spacings on levels $l>0$ are related to that on level 0 by $\Delta x_l = \Delta x_0 / 2^l$. The process of advancing the computation through these levels is recursive. We first advance the grids on level 0 by a time $\Delta t_0$, and then we advance the grids on level 1 by two timesteps of size $\Delta t_1 = \Delta t_0/2$. However each of these advances is followed by two advances on level 2, and so forth, such each level 0 advance involves $2^l$ advances of timestep $\Delta t_0/2^l$ on level $l$. At the end of each advance of level $l>0$ through 2 steps, we perform a synchronization procedures between levels $l$ and $l-1$ to ensure conservation of mass, momentum, and energy across the interface between the two levels. We set the overall timestep $\Delta t_0$ by computing the Courant condition (including a contribution to the effective signal speed from radiation pressure -- \citealt{krumholz07b}) separately on each level at the beginning of each coarse timestep. We then set $\Delta t_0 = \min(2^l \Delta t^l)$, ensuring that each level obeys the Courant condition for its advances.

\subsection{Refinement and Boundary Conditions}

The ORION AMR framework automatically adds and removes higher resolution grids throughout a simulation. We determine when higher resolution grids are required based on the following criteria:
\begin{enumerate}
\item Any cell with a density greater than half the initial density at the edge of the cloud (see below) must be refined to at least level 1.
\item We refine any cell within whose distance $d$ to the nearest star particles is less than $16\Delta x_l$. This ensures that regions around stars are always refined to the maximum level.
\item We refine any cell where the density exceeds the Jeans density, given by
\begin{equation}
\rho_J = J^2 \frac{\pi c_s^2}{G \Delta x_l^2},
\end{equation}
where $c_s$ is the sound speed and we use $J=1/16$. This avoids artificial fragmentation \citep{truelove97}.
\item We refine any cell where the gradient in the radiation energy density satisfies
\begin{equation}
|\nabla E| > 0.15 \frac{E}{\Delta x_l}.
\end{equation}
This ensures that we adequately resolve gradients in the radiation energy density.
\end{enumerate}
These conditions are applied recursively to every level up to some pre-specified maximum level $L$.  In practice, the fourth condition is usually the most stringent.

\begin{deluxetable*}{cccccccccccc}
\tablecaption{Simulation Parameters\label{runsetup}}
\tablehead{
\colhead{Name} &
\colhead{$M$ $(\msun)$} &
\colhead{$\Sigma$ (g cm$^{-2}$)} & 
\colhead{$R$ (pc)} &
\colhead{$\sigma_v$ (km s$^{-1}$)} &
\colhead{$t_{\rm ff}$ (kyr)} &
\colhead{$L_{\rm box}$ (pc)} &
\colhead{$L$} &
\colhead{$N_0$} &
\colhead{$\Delta x_0$ (AU)} &
\colhead{$N_L$} & 
\colhead{$\Delta x_L$ (AU)}
}
\startdata
L & 100 & 0.1 & 0.258 & 1.29 & 217 & 1.95 & 8 & 256 & 1573 & 65,536 & 6.14 \\
M & 100 & 1.0 & 0.081 & 2.30 & 38.6 & 0.489 & 6 & 256 & 393 & 16,384 & 6.14 \\
H & 100 & 10.0 & 0.026 & 4.08 & 6.86 & 0.160 & 5 & 168 & 197 & 5,376 & 6.14 \\
\enddata
\tablecomments{Col.\ 7: linear size of computational domain. Col.\ 8: maximum refinement level. Col.\ 9: number of cells per linear dimension on the coarsest level. Col.\ 10: linear cell size on the coarsest level. Col.\ 11-12: same as col.\ 9-10, but for the finest level.
}
\end{deluxetable*}

For all our calculations we use symmetry boundary conditions on the hydrodynamics, but we use adaptivity to remove the boundary far enough from the cloud we are simulating so that no part of it ever approaches the boundary. The gravity module uses Dirichlet boundary conditions on the potential, with the potential on the boundary set equal to the value obtained from an octopole expansion of the density distribution inside the computational domain. The radiation module uses Marshak boundary conditions, meaning that radiation energy generated within the domain is free to escape. The incoming radiation flux is set to that appropriate for a blackbody radiation field at a temperature of 20 K.

\subsection{Initial Conditions and Simulation Setup}

The initial setup for all of our simulations is similar to those of \citetalias{krumholz07a}. In all cases the setup consists of an initially spherical cloud of mass $M=100$ $\msun$, initial mean surface density $\Sigma$, and radius  $R = \sqrt{M/(\pi \Sigma)}$ in the center of a cubical computational domain. Observed dense star-forming clumps of molecular gas have roughly powerlaw density structures $\rho\propto r^{-\krho}$ with $\krho \simeq 1.5$ and considerable scatter \citep{caselli95a,  beuther02c, beuther05b, beuther06a, mueller02, sridharan05}, so, following \citet{mckee03a}, we adopt a powerlaw density structure with $\krho=1.5$ for our initial conditions. We give our clouds an initial turbulent velocity field chosen to put them in approximate balance between gravity and turbulent ram pressure. The velocity dispersion is
\begin{equation}
\sigma_v = \sqrt{\frac{GM}{2(\krho-1) R}},
\end{equation}
and the initial velocity field we use is identical to that of run 100A of \citetalias{krumholz07a}, which is a Gaussian-random field with a power spectral density $P(k) \propto k^{-2}$. Finally, we give the gas in the cloud an initial temperature $T_g = 20$ K, and we set the initial radiation energy density throughout the computational domain to $E = 1.2\times 10^{-9}$ erg cm$^{-3}$, the value for a blackbody radiation field at a temperature $T_r = 20$ K.  

Outside the cloud we place a hot, diffuse ambient medium with a density $\rho_a = \rho_{\rm edge}/100$, where
\begin{equation}
\rho_{\rm edge} = \left(\frac{3-\krho}{4\pi}\right) \frac{M}{R^3}
\end{equation}
is the density at the cloud edge. The ambient medium has a temperature $T_a = 100 T_g$, ensuring that it is in thermal pressure balance with the cloud. To ensure that the ambient medium does not cool, radiatively heat the cloud, or interfere with radiation escaping from the cloud, we set the opacity of the ambient medium to a numerically small value. 

We simulate three different clouds, chosen with values of $\Sigma$ to be representative of three different types of star-forming environment as discussed in Section \ref{intro}. The first run has $\Sigma=0.1$ g cm$^{-2}$, typical of diffuse star-forming clouds such as Perseus and Ophiuchus. The second has $\Sigma=1.0$ g cm$^{-2}$, typical of regions of massive star formation in the Galaxy. The third has $\Sigma=10$ g cm$^{-2}$, an extremely high surface density found only in clusters near the Galactic center and in extragalactic super star clusters. However, this type of star-forming environment is believed to have been more common earlier in cosmological evolution. We call the runs L, M, and H, for low, medium, and high column density. We summarize the setup of the three runs in Table \ref{runsetup}. We run each simulation for a time $t=0.6\tff$, where
\begin{equation}
\tff = \sqrt{\frac{3\pi}{32G\overline{\rho}}}
\end{equation}
is the free-fall time computed at the mean density $\overline{\rho} = 3M/(4\pi R^3)$ of the cloud. By this point the differences between the runs are clearly established, and the fraction of the collapsed mass in the most massive star asymptotes to a constant value (see below).

We emphasize that we have chosen the numerical setup so that the runs are, as much as possible, simply rescaled versions of one another. The initial conditions have identical density structures, virial ratios, and velocity fields, and the refinement criteria and peak resolution is the same in every run. We simulate each cloud for the same number of free-fall times. The homology between the runs is broken only by the influence of radiation. Radiation fixes the gas temperature (and thus causes the initial Mach numbers of the runs to vary slightly), and, much more importantly, the very different optical depths of the different clouds cause them to respond differently once stellar feedback begins. Thus we expect any difference between the three runs to be almost entirely dictated by their differing response to stellar radiative forcing.

\section{Results}
\label{results}

\subsection{Collapse Morphology}

\begin{figure}
\epsscale{1.17}
\plotone{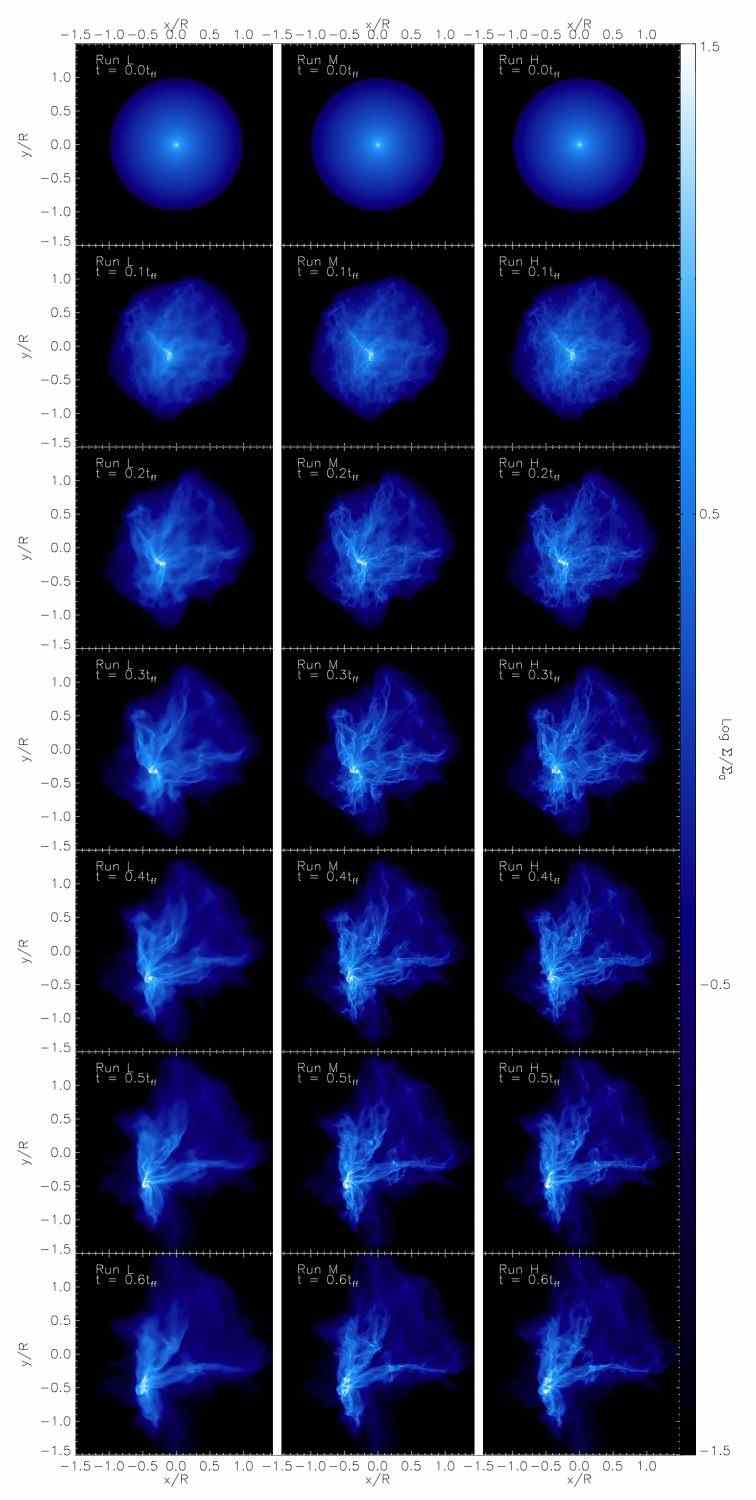}
\epsscale{1.0}
\caption{
\label{dencomp_large}
Column density in simulations L, M, and H (\textit{left to right column}) at times running from $t=0$ to $t=0.6\tff$ (\textit{top to bottom row}). The color scale is normalized to the initial mean column density $\Sigma_0=0.1$, $1$, and $10$ g cm$^{-2}$ for runs L, M, and H, respectively.
}
\end{figure}

We show the large-scale evolution of runs L, M, and H in Figure \ref{dencomp_large}. As the plot shows, and as expected, the three runs are essentially homologous on large scales. Once we account for the scaling of cloud radius and surface density between the runs, the only noticeable difference is a slight trend toward increasingly filamentary structures as $\Sigma$ increases. This is easily understood as a radiative effect. As noted above, runs with higher $\Sigma$ require larger velocity dispersions $\sigma_v$ to maintain initial virial balance, while the sound speed is fixed by radiative effects. Thus the Mach number of the initial turbulence increases with $\Sigma$, and this produces the observed increase in filamentary structure.

\begin{figure*}
\epsscale{1.17}
\plottwo{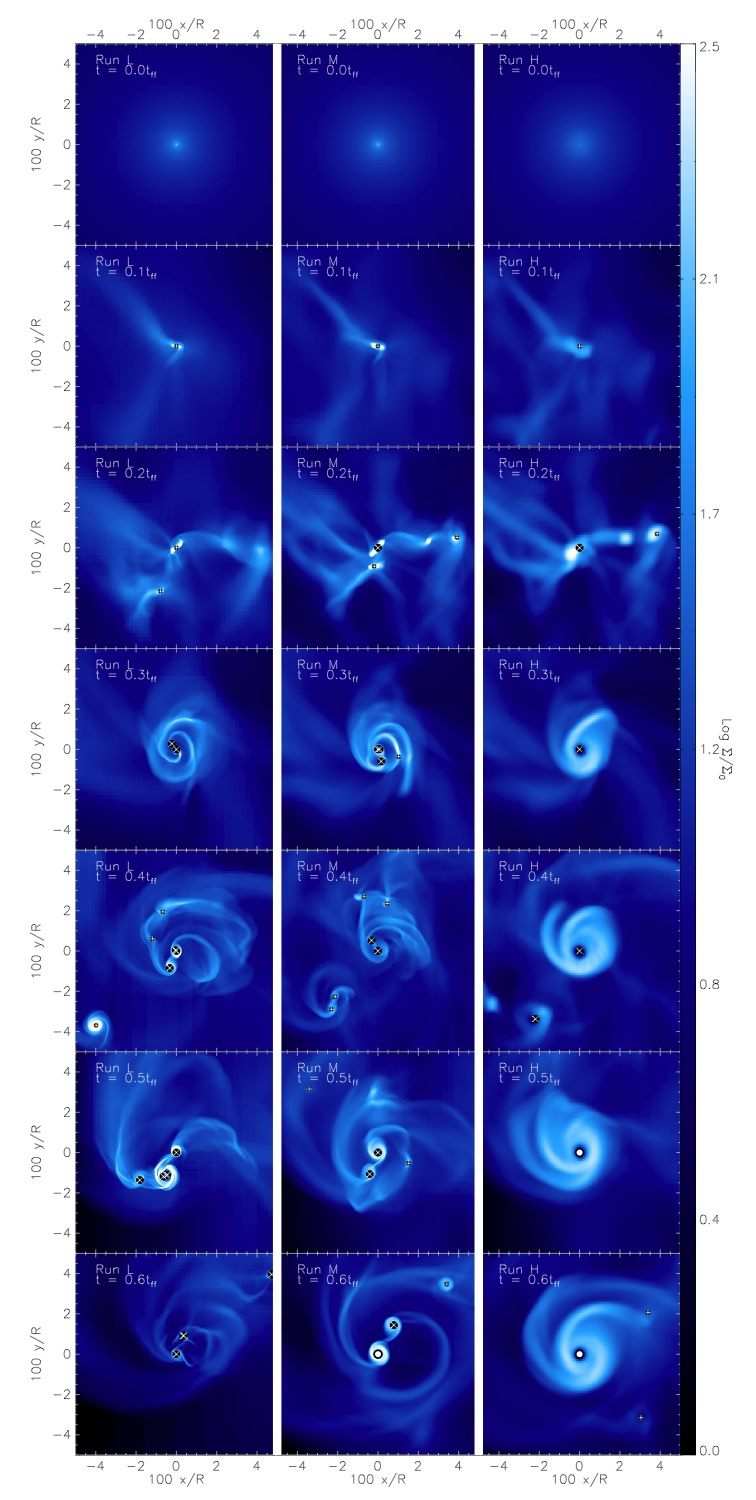}{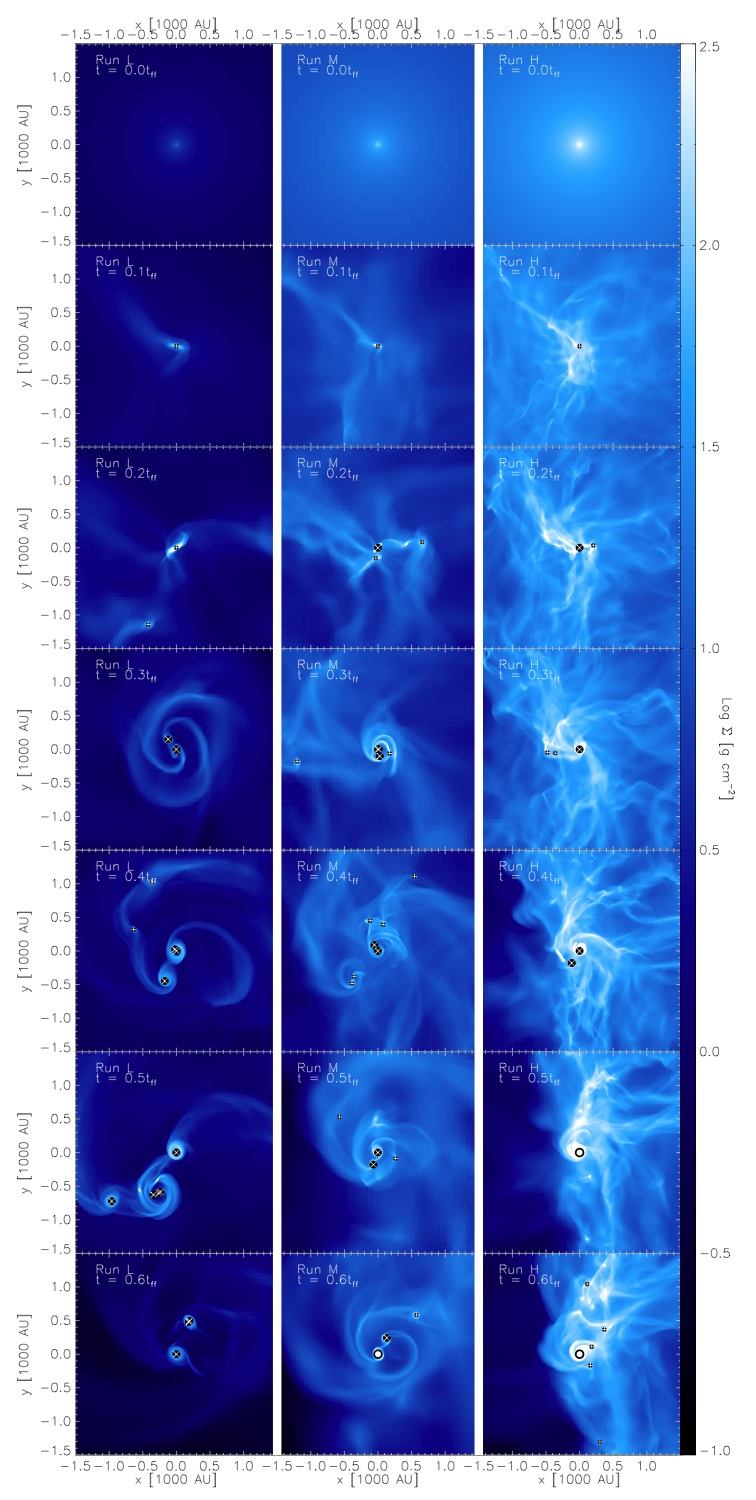}
\epsscale{1.0}
\caption{
\label{dencomp_small}
Column density in simulations L, M, and H (\textit{left to right column}) at times running from $t=0$ to $t=0.6\tff$ (\textit{top to bottom row}). Symbols indicate stars, with the type of symbol indicating the stellar mass. Low mass stars ($M_* = 0.05-1$ $\msun$) are indicated by $+$ signs, intermediate mass stars ($M_*=1-8$ $\msun$) by $\times$ signs, and massive stars ($M_*>8$ $\msun$) by filled circles. {\it Left:} the region shown is a $0.1R \times 0.1R$ box centered on the most massive star, or the origin if no stars are present, and the color scale is normalized to the initial mean column density $\Sigma$, where $R$ and $\Sigma$ have the values given in Table \ref{runsetup} for runs L, M, and H. {\it Right:} the region shown is a $3000\mbox{ AU} \times 3000\mbox{ AU}$ box centered on the same point as in the left panel, and the color scale is in the same physical units for every run. 
}
\end{figure*}

In Figure \ref{dencomp_small}, we show the simulations at the same times as in Figure \ref{dencomp_large}, but now zoomed in to a small region centered on the most massive star present, or on the origin before stars form. In the left panel of Figure \ref{dencomp_small} the images are scaled homologously, so that the regions shown all have a length equal to 10\% of the initial cloud radius, and the color scale is in units of surface density divided by initial mean surface density. In the right panel the scaling is physical, so that all the plots show a region of fixed physical size, using a column density scale in fixed rather than normalized units.

In contrast to the large-scale homology seen in Figure \ref{dencomp_large}, on small scales the runs rapidly diverge. Deviations from homology start to appear around $0.2\tff$, and by $0.4\tff$ any visual similarity between the runs is completely gone. Progressing from low to high $\Sigma$, runs are characterized by increasing surface densities even when normalized to the initial surface density. In run H the predominant structure is a single large disk concentrated around a single central object. In run M we have a massive binary with two circumstellar disks and a larger circumbinary disk. Finally, in run L the disks are much smaller and less dense, and they have mostly depleted by the final time shown.

\subsection{Fragmentation and Star Formation}

\begin{deluxetable*}{cccccccccccccccc}
\tablecaption{Stellar Content versus Time\label{starcontent}}
\tablecolumns{10}
\tablehead{
\colhead{$t/\tff$} &
&
\colhead{$N_*$} &
\colhead{$M_{*,\rm tot}$} &
\colhead{$M_{*,\rm max}$} &
\colhead{$f_{\rm max}$} &
&
\colhead{$N_*$} &
\colhead{$M_{*,\rm tot}$} &
\colhead{$M_{*,\rm max}$} &
\colhead{$f_{\rm max}$} &
&
\colhead{$N_*$} &
\colhead{$M_{*,\rm tot}$} &
\colhead{$M_{*,\rm max}$} &
\colhead{$f_{\rm max}$} 
}
\startdata
& \qquad\qquad &
\multicolumn{4}{c}{Run L} &
\qquad\qquad &
\multicolumn{4}{c}{Run M} &
\qquad\qquad &
\multicolumn{4}{c}{Run H} \\
0.0 & &  0 & \nodata & \nodata & \nodata  & &  0 & \nodata & \nodata & \nodata  & &  0 & \nodata & \nodata & \nodata  \\
0.1 & &  1 &  0.18 &  0.18 &  1.00 & &  1 &  0.29 &  0.29 &  1.00 & &  1 &  0.37 &  0.37 &  1.00 \\
0.2 & &  2 &  0.95 &  0.86 &  0.90 & &  3 &  1.55 &  1.16 &  0.75 & &  2 &  1.91 &  1.86 &  0.97 \\
0.3 & &  3 &  5.06 &  2.54 &  0.50 & &  6 &  5.73 &  2.56 &  0.45 & &  3 &  5.60 &  5.13 &  0.92 \\
0.4 & &  7 &  8.10 &  2.93 &  0.36 & &  7 &  9.61 &  5.07 &  0.53 & &  2 &  8.16 &  7.10 &  0.87 \\
0.5 & &  7 & 11.55 &  4.06 &  0.35 & &  4 & 12.44 &  7.18 &  0.58 & &  2 & 10.92 & 10.83 &  0.99 \\
0.6 & &  8 & 15.58 &  5.75 &  0.37 & &  7 & 16.42 &  8.77 &  0.53 & &  9 & 14.96 & 13.41 &  0.90 \\
\enddata
\tablecomments{Col.\ 1: run time. Col.\ 2: number of stars present in run L. Col.\ 3: total mass of stars in run L. Col.\ 4: mass of largest star in run L. Col.\ 5: fraction of total stellar mass in largest star. Col.\ 6-9 and 10-13: same as columns 2-5, but for runs M and H.
}
\end{deluxetable*}

\begin{figure}
\plotone{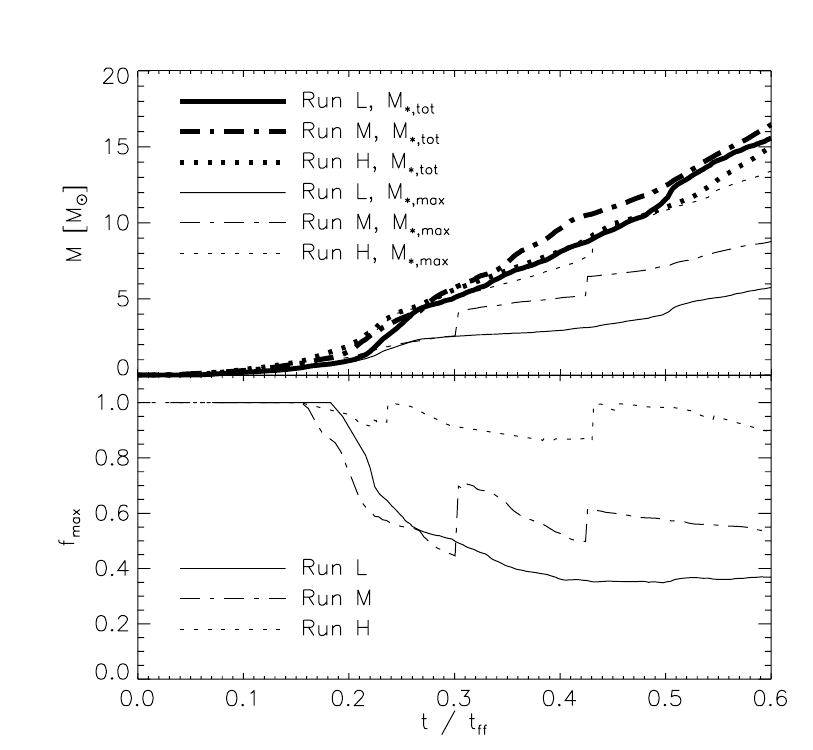}
\caption{
\label{starhist}
Star formation histories in runs L, M, and H. {\it Top:} total stellar mass $M_{*,\rm tot}$ ({\it thick lines}) and mass of most massive star $M_{*,\rm max}$ ({\it thin lines}) versus time, as indicated. {\it Bottom:} fraction $f_{\rm max}$ of total stellar mass in most massive star versus time. Sharp jumps represent mergers between a central massive star and a smaller star.
}
\end{figure}

\begin{figure}
\plotone{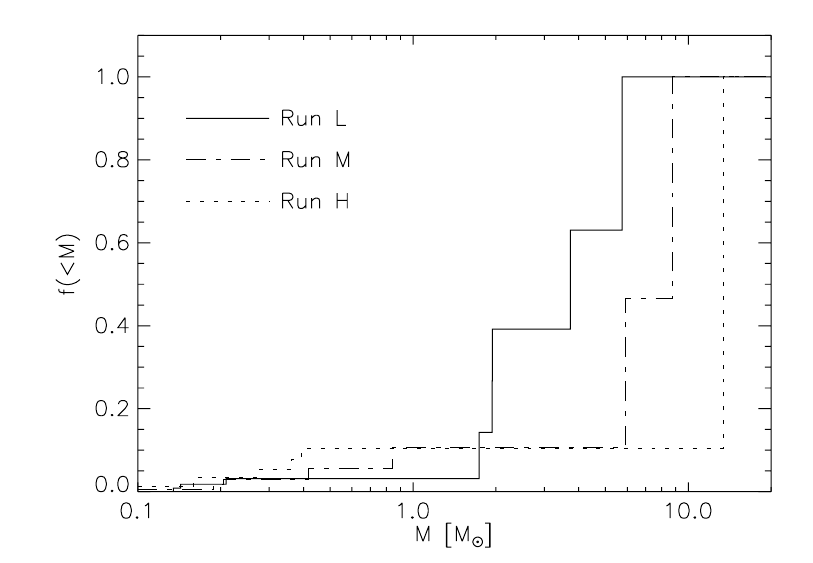}
\caption{
\label{cumdist}
Fraction $f(<M)$ of total stellar mass contained in stars with mass $<M$ as a function of $M$, for each of the runs at time $t=0.6\tff$.
}
\end{figure}

The higher $\Sigma$ runs also fragment less, producing fewer, more massive stars than the low $\Sigma$ runs. We show this in Table \ref{starcontent} and Figure \ref{starhist}. The total mass in stars $M_{*,\rm tot}$ at any given time (normalized to $\tff$) is very similar from run to run, varying by less than 10\% from run L to run H at times $>0.2\tff$. At the end of the simulation, each run has a total of 15 $\msun$ of stars. This reflects that the star formation rate in the simulations is driven by large-scale flows that are changed very little by radiation feedback. However, the pattern of fragmentation is quite different. The mass of the most massive object $M_{*,\rm max}$ in the three runs begins to diverge at $0.2-0.3\tff$, and thereafter it increases much more rapidly in run H than in run L. At the final time the difference in mass is nearly a factor of 3, with run L having a maximum stellar mass below 6 $\msun$, and run H reaching 15 $\msun$. In all the runs the fraction of the total stellar mass in the most massive object $f_{\rm max}$ asymptotes to a roughly constant value after $0.3\tff$. The asymptotic value ranges from $f_{\rm max}\sim 0.35$ in run L to $f_{\rm max}\sim 0.9$ in run H. In effect the initial gas cloud in run H is like a single massive protostellar core that forms one massive star plus a few small secondaries, while the cloud in run L instead forms a small cluster that does not include any massive stars. Run M is intermediate.

The difference in runs is even more apparent if we focus on a single time. Figure \ref{cumdist} shows the cumulative distribution function of stellar mass at $t=0.6\tff$ in each of the runs. In run $L$ the distribution function rises relatively smoothly between 1 and 5 $\msun$, so the system consists of a number of stars of roughly comparable mass. In contrast, in run M 90\% of the mass is in just two stars that form a binary system, a result very similar to that in \citetalias{krumholz07a}, which used an initial surface density $\Sigma=0.7$ g cm$^{-2}$. In run H a comparable fraction of the mass is in a single star. Since the system in run L involves a number of stars of comparable mass, these are unlikely to wind up as a bound star system. The system in run M, on the other hand, is stable and is likely to remain a bound binary. Thus in both runs M and H, the result is that most of the mass goes into a single star system, while in run L the mass will end up divided into several star systems.

\subsection{Thermal Structure}

\begin{figure*}
\epsscale{1.17}
\plottwo{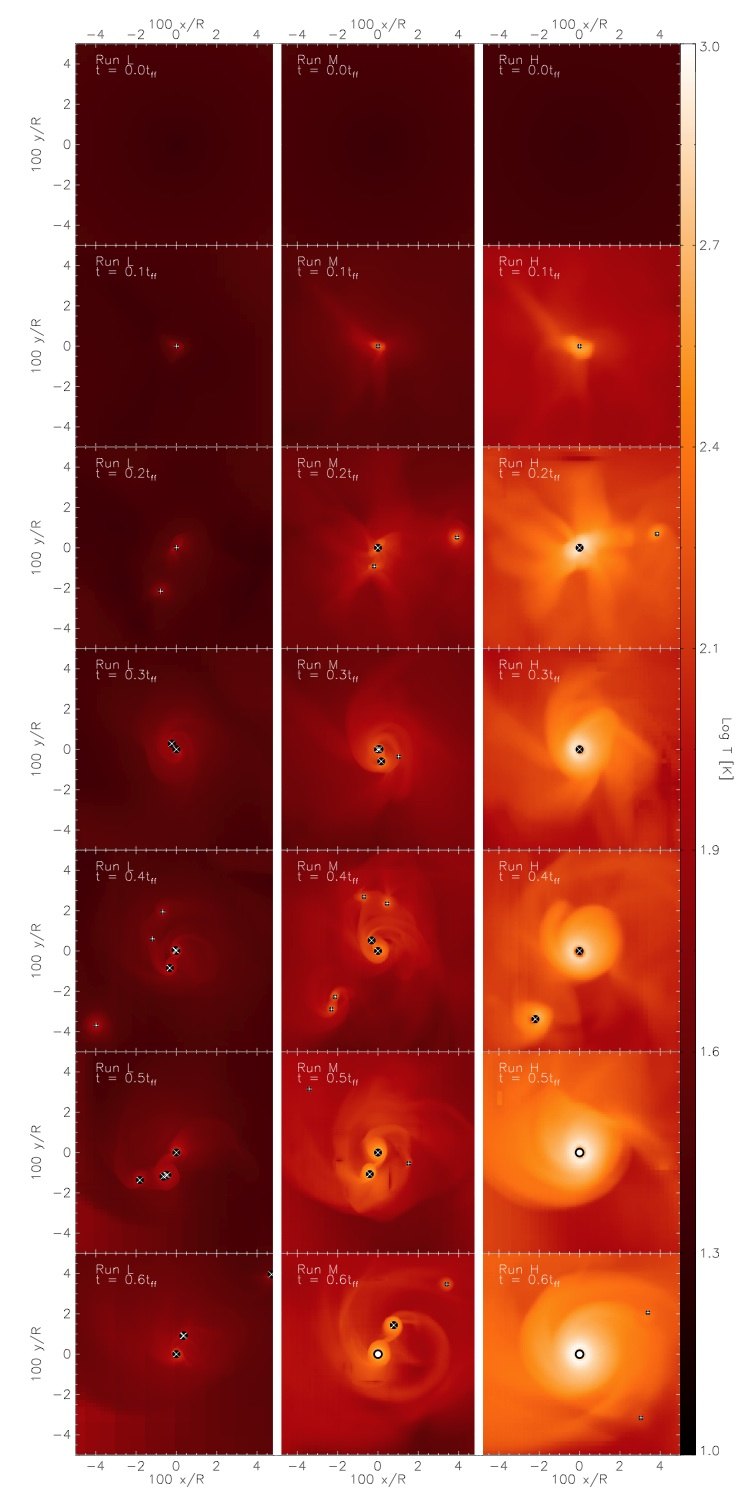}{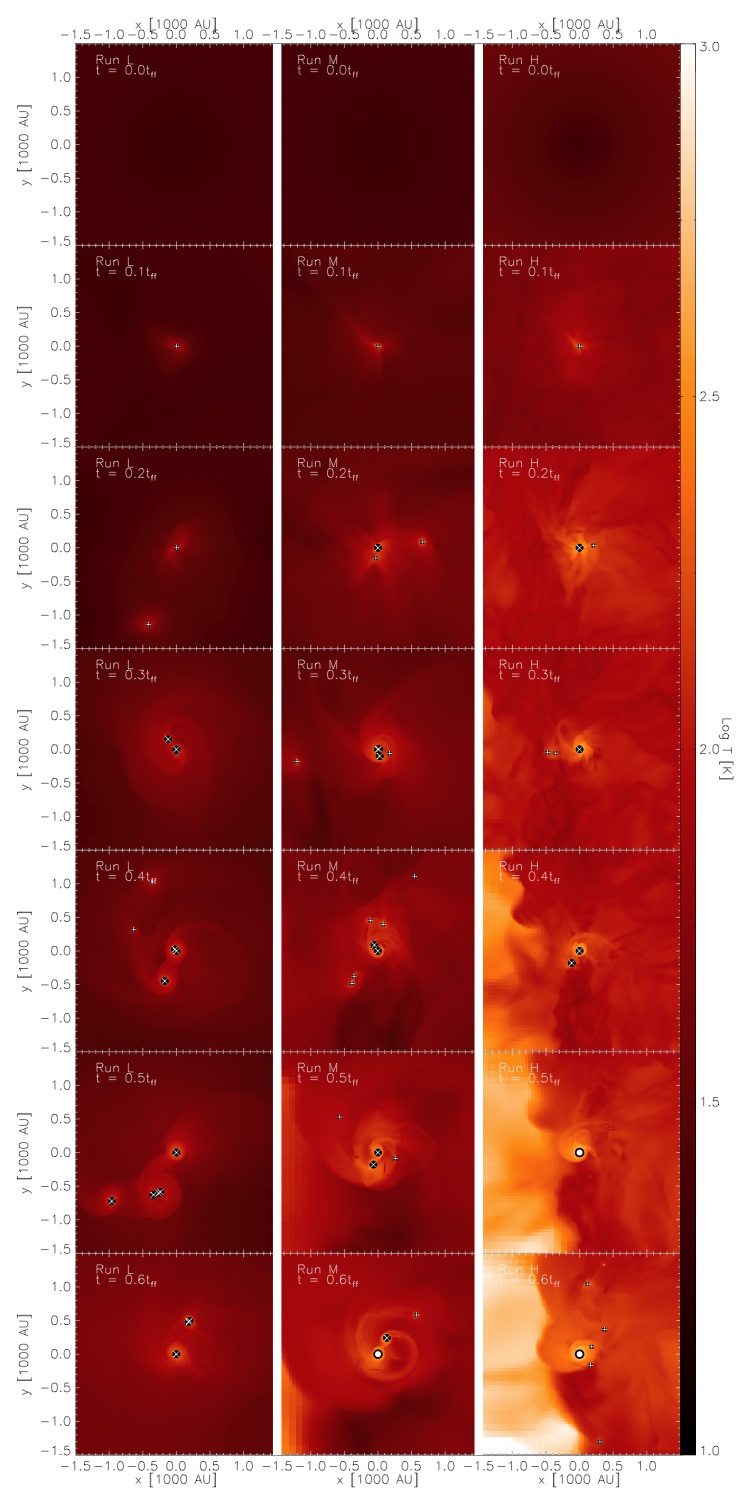}
\epsscale{1.0}
\caption{
\label{tempcomp_small}
Same as Figure \ref{dencomp_small}, except that the plots show column density-weighted temperature, defined as $\int \rho T\, dz / \int \rho\, dz$. The color scales are the same in both the left and right sides.
}
\end{figure*}

The difference in morphology, fragmentation, and star formation is easy to understand if we examine the temperature structure of the gas. In Figure \ref{tempcomp_small} we show the column-density weighted temperature over the same small-scale regions as in Figure \ref{dencomp_small}. Clearly the temperature distribution in the gas in the runs is even less homologous than the density structure. Moreover, the differences in temperature begin to appear at earlier times. At $t=0.1\tff$ the density plots are nearly indistinguishable once the homologous scaling is removed (left panel of Figure \ref{dencomp_small}), while the differences in temperature are already obvious. It is important to point out that at this point there no stars present that are producing significant amounts of power via nuclear luminosity. The most massive star present in any of the runs is only $0.37$ $\msun$, and its luminosity is entirely driven by accretion. The difference in temperature is therefore solely due to the two factors pointed out by \citetalias{krumholz08a}: compared to run L, run H has both a higher density to produce higher accretion rates and thus higher accretion luminosities, and a higher optical depth to more effectively trap the radiation that is produced.

\begin{figure}
\epsscale{1.17}
\plotone{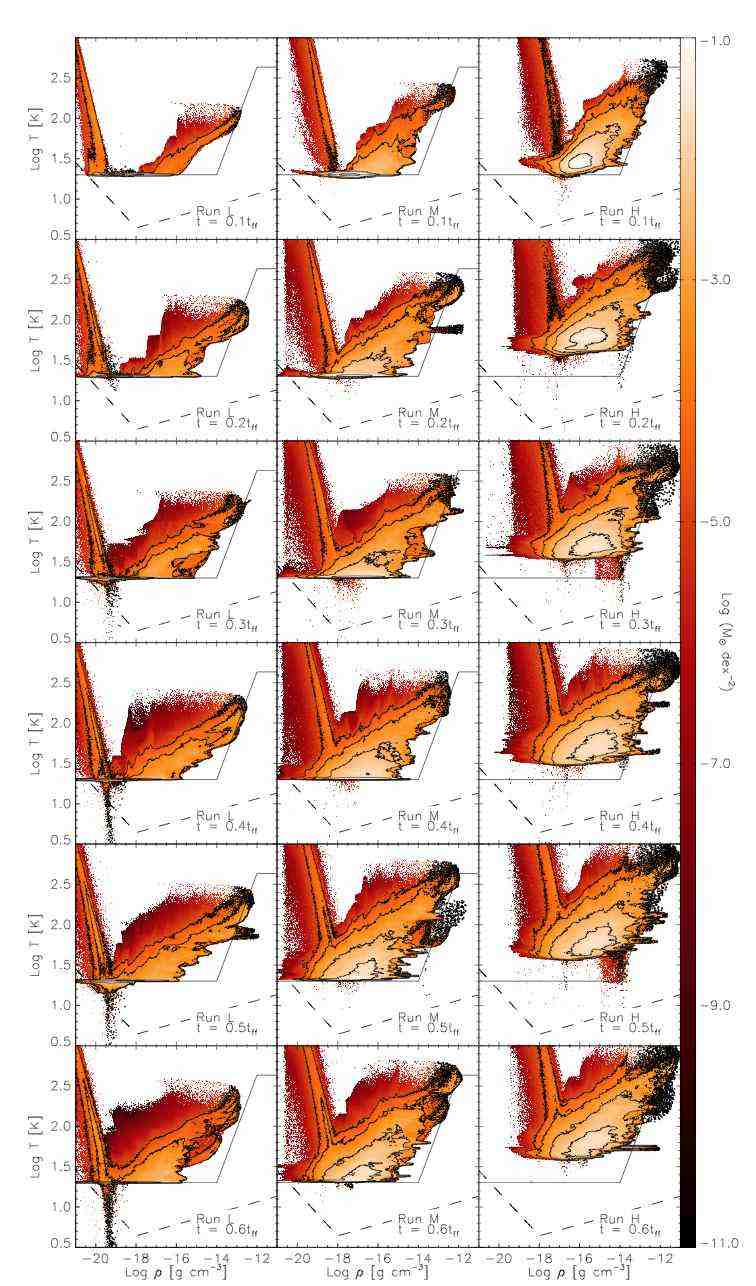}
\epsscale{1.0}
\caption{
\label{rhotplane}
Temperature-density relation for runs L, M, and H (left to right columns) at the same times as in Figure \ref{dencomp_large}, except that we omit time $t=0.0$. Colors show the mass density per square dex in the $\log \rho - \log T$ plane. Contours from outermost to innermost indicate the region in the $\log\rho-\log T$ plane containing 99.9\%, 99\%, 90\%, and 50\% of the total gas mass at that time (i.e.\ not including the mass in stars). In some runs only the outermost contours are visible because the inner ones form a thin line near $T=20$ K. The linear feature extending to high temperature and low density represents cells that contain a mix of cloud and hot ambient medium; such cells never contain a significant fraction of the cloud mass, as indicated by the contours. The solid line is the barotropic curve of \citet{dobbs05}, which has a constant temperature $T=20$ K at low density. The dashed line is the optically thin cooling approximation of \citet{larson05}.
}
\end{figure}

We can understand the difference in the thermal structure of the runs more quantitatively, and explore how this difference is likely to influence fragmentation, by examining the relationship between temperature and density in each of the runs. Figure \ref{rhotplane} shows the locus occupied by the clouds in runs L, M, and H at different times in the simulation in the plane of log density versus log temperature. The color represents the mass density at a given point, and the contours indicate, from lowest to highest, regions in the plane containing 99.9\%, 99\%, 90\%, and 50\% of the gas mass in the computational domain.

The figure demonstrates how different the thermal structure of the gas is in each of the runs. In run L, the great majority of the gas is near the background temperature of 20 K at all times. Even at the final time only 10\% of the gas mass is at noticeably elevated temperatures, and less than 1\% of the mass is heated above 100 K. In contrast, in run H the heating is much more extensive. Even at $t=0.1\tff$, when the only source of heating is the accretion luminosity of a $0.37$ $\msun$ star, the contours containing 50\% and 90\% of the mass are noticeably elevated above the $T=20$ K line. Deuterium burning in the star begins shortly before $0.2\tff$, and by the final time deuterium burning and Kelvin-Helmholtz contraction (the star has not yet reached the main sequence) provide enough luminosity to keep all the mass at temperatures $\ga 50$ K. Run M is intermediate, with small but significant fractions of the cloud mass reaching elevated temperatures at early times, and more mass becoming heated as the run progresses and the stars become more luminous. This is consistent with the analytic predictions of \citetalias{krumholz08a}, who find that a column density of $1$ g cm$^{-2}$ constitutes the rough line between clouds that do and do not experience significantly elevated temperatures over much of their mass as a result of trapped accretion luminosity.

It is important to point out that the temperature does not need to rise to the point where the Jeans mass is above 100 $\msun$ in order to inhibit fragmentation. Indeed, such a rise in temperature would be sufficient to halt collapse of the core entirely. Instead, the heating prevents fragmentation by creating an environment where the effective equation of state with $\gamma = 1 + d\log T/d\log \rho > 1$ throughout the bulk of the cloud mass. Examining Figure \ref{rhotplane}, we see that the region containing 90\% of the cloud mass (the third contour from the outermost one) is almost perfectly horizontal in run L at all times, so $\gamma\approx 1$. In runs M and H, on the other hand, this region has a slope $\sim 0.2-0.3$ in the $\log \rho-\log T$ plane at all times of $0.2\tff$ or more, similar to the result obtained by \citetalias{krumholz07a}, indicating that the effective equation of state is closer to $\gamma=1.2-1.3$. As \citet{larson05} points out, and the simulations of \citet{jappsen05} confirm, fragmentation is likely as long as the effective equation of state for the gas is $\gamma \leq 1$, and is unlikely when $\gamma > 1$. In runs M and H, there is no significant gas mass with $\gamma \leq 1$, which is why fragmentation is suppressed.

Finally, we emphasize that these phenomena cannot be correctly captured by analytic equations of state that are based on either a baroptropic or and optically thin cooling assumption. Examples of such equations of state from \citet{dobbs05} and \citet{larson05} are shown in Figure \ref{rhotplane}, and they clearly do not even come close to reproducing the results with radiative transfer, a point also made by \citet{boss00}, \citet{krumholz06b}, \citet{krumholz07a}, and \citet{offner09a}. Any such approximation would give the same temperature-density relation for all three of our simulated clouds, while clearly the results are different at different times and for different initial cloud column densities. \citet{urban10a} reach the same conclusion for lower-density, larger-scale clouds based on their simulations.

Although we have not tested the \citet{bate09a} approach of omitting radiation from stars and including only radiative emission by gas on size scales resolved by the computation (which are much larger than stellar scales in both Bate's calculation and ours), it seems unlikely that this approximation could succeed either in the case of clouds with differing initial column densities. It would capture the difference in optical depth between runs, but it would not capture the effect that higher density runs produce higher accretion rates and thus higher accretion luminosities from the embedded protostars. The analytic models of \citetalias{krumholz08a} suggest that both effects are of comparable importance.

\section{Discussion}
\label{discussion}

\subsection{The Massive Star Fraction}

Our results demonstrate that the amount of fragmentation that a cloud undergoes is likely to depend strongly on its surface density, and this has important implications for where we expect massive stars to form. Consider a protostellar core, an object with a mass of a few tenths to a few hundreds of $\msun$ that collapses to make one or more stars. Low mass cores do not have significant internal turbulence \citep{andre07a, kirk07a, rosolowsky08a}, as is expected on theoretical grounds \citep{offner08a}. Consequently, while they may fragment into a binary like run M, we expect most of the stellar mass they produce to end up in a single star system. The overall efficiency of turning gas mass into stellar mass is expected to be $\epsilon \approx 1/3$ rather than $\epsilon = 1$ as a result of mass ejection by protostellar outflows \citep{matzner00, alves07, enoch08a}.

\begin{figure}
\plotone{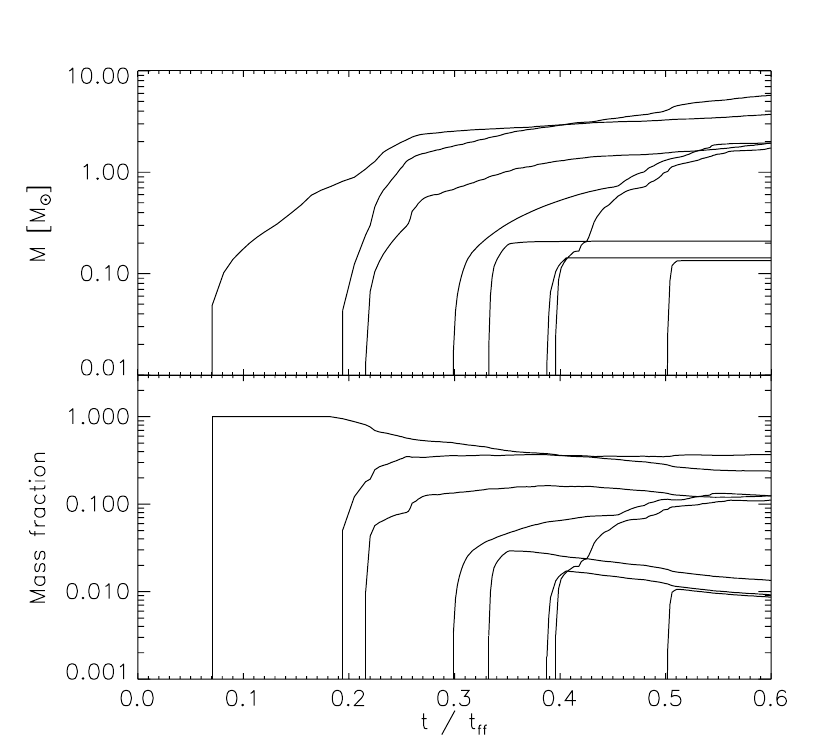}
\caption{
\label{runLfrag}
Mass (top panel) and fraction of the total stellar mass (bottom panel) in all stars as a function of time in run L. Each line represents an individual star.
}
\end{figure}

In contrast, as pointed out by \citet{mckee03a}, massive cores such as those we simulate are turbulent, since their masses are many times the thermal Jeans mass. We find that, at low $\Sigma$, such cores will fragment so that their stellar mass is divided among many star systems. Figure \ref{runLfrag} gives a more detailed picture of how this happens in run L. There are five fragments whose masses appear to be asymptotically approaching fixed fractions of the total stellar mass, ranging from 11\% to 37\%, plus three more smaller stars whose masses appear to have reached nearly fixed maxima, and that are therefore declining with time in the total fraction of stellar mass that they represent.

We can use this result to make a toy model for how the fraction of the stellar mass that is in massive stars is likely to vary with surface density. We begin from the observation that stars form from cores that have a mass distribution with the same functional form as the IMF, so that the IMF is set at the phase when gas fragments into protostellar cores \citep{motte98, testi98, johnstone01, onishi02, beuther04, reid05, reid06a, reid06b, alves07, nutter07a, simpson08a, enoch08a, rathborne09a}. We model this core mass function (CMF) using a \citet{chabrier05a} stellar system IMF shifted to higher mass by a factor of 3 to account for the mass that is ejected by outflows:
\begin{equation}
\frac{dn_c}{d\ln m_c} = 
\left\{
\begin{array}{ll}
A \exp\left[-\frac{(\ln m_c - \ln \overline{m_c})^2}{2\sigma^2}\right], \quad & m_c < m_{\rm break} \\
B \left(\frac{m_c}{m_{\rm break}}\right)^{-1.3}, & m_c \geq m_{\rm break},
\end{array}
\right.
\end{equation}
with $\overline{m_c} = 0.75 \msun$, $m_{\rm break} = 3\msun$, and $\sigma = 0.55$. Our values of $\overline{m_c}$ and $m_{\rm break}$ are chosen so that, when $\epsilon=1/3$, the stellar IMF will have a peak at $0.25$ $\msun$ and will break from lognormal to powerlaw form at $1.0$ $\msun$, in agreement with \citeauthor{chabrier05a}'s best fit to observations.
The normalization factors are related by $B = A \exp[-(\ln m_{\rm break} - \ln \overline{m_c})^2/(2\sigma^2)]$. Theoretical models are able to explain this distribution of core masses as arising naturally from the properties of supersonic turbulence \citep{padoan02, hennebelle08b}. In this picture, the final stellar IMF is simply the convolution of the core mass function (CMF) with a simple function that maps core mass to stellar mass, and which must be nearly mass-independent. \citet{clark07} suggest that this correspondence will be disrupted if the core free-fall time is mass-dependent. However, observed cores do not have mass-dependent free-fall times \citep{andre07a}, and theoretical models predict that, contrary to \citeauthor{clark07}'s assumption, core free-fall time depends on mass at most very weakly \citep{mckee03a, hennebelle09a} as a result of turbulent support.
 
To make a quantitative model of how fragmentation in low $\Sigma$ regions will affect the IMF, we consider two extreme scenarios that bracket the outcome of run L. The first is that, in regions of low $\Sigma$, cores with mass $m_c$ above some minimum fragmentation mass $m_{\rm frag}$ produce only a single massive star with a mass $m_* = m_c/9$, i.e.\ 2/3 of the core mass is ejected by outflows, 1/3 of what remains goes into the largest star, and the remaining 2/3 goes into low mass stars with $m_* < m_{\rm frag}/3$. The alternative possibility is that the cores with mass $m_c>m_{\rm frag}$ fragment into $n_{\rm frag}$ stars of equal mass $m_* = m_c/(3n_{\rm frag})$, with the factor of 3 to account for ejection by outflows. Based on the outcome in run L, we adopt $n_{\rm frag} = 5$ for our toy model, although of course in reality we expect that the number of fragments and their mass distribution will vary stochastically.

In the first scenario, if the total mass of cores with masses between $m_c$ and $m_c+dm_c$ is given by $dn_c/d\ln m_c$, the corresponding mass of stars with masses between $m_*=m_c/3$ and $m_* + dm_*$ is given by $dn_*/d\ln m_* = (1/9) dn_c/d\ln m_c$ for any mass $m_* > m_{\rm frag}$. In the second scenario, we instead have $m_* = m_c/(3 n_{\rm frag})$ and $dn_*/d\ln m_* = (1/3) dn_c/d\ln m_c$, since all of the core mass that is not ejected by outflows (i.e.\ 1/3 of it) goes into stars of mass $m_c/(3 n_{\rm frag})$. Finally, in high $\Sigma$ regions where massive cores do not fragment, we also have $dn_*/d\ln m_* = (1/3) dn_c/d\ln m_c$, but now the stellar mass is related to the core mass by $m_* = m_c/3$ rather than $m_* = m_c/(3 n_{\rm frag})$.

Given these relations, we can compute the fraction of all stellar mass that is contained in stars with masses greater than $m_*$, which we denote $f(>m_*)$. We adopt a minimum stellar mass $m_{*,\rm min} = 0.01$ $\msun$, and a maximum $m_{*,\rm max} = 120$ $\msun$ \citep{figer05}. First consider regions of high $\Sigma$ where massive cores do not fragment and stellar and core masses are related by $m_* = m_c/3$. In such a region, we have
\begin{equation}
f(>m_*) = \frac{
\int_{3 m_*}^{3 m_{*,\rm max}} \left(\frac{1}{3}\right)\frac{dn_c}{d\ln m_c} \, dm_c
}{
\int_{3 m_{*,\rm min}}^{3 m_{*,\rm max}} \left(\frac{1}{3}\right)\frac{dn_c}{d\ln m_c} \, dm_c
}
\equiv f_H(>m_*).
\label{fh}
\end{equation}
This equation simply states that the fraction of stellar mass in stars larger than $m_*$ is the same as the fraction of core mass in cores larger than $3 m_*$. For low $\Sigma$ regions, in the first scenario we instead have
\begin{equation}
f(>m_*) = \frac{
\int_{9 m_*}^{9 m_{*,\rm max}} \left(\frac{1}{9}\right)\frac{dn_c}{d\ln m_c} \, dm_c
}{
\int_{3 m_{*,\rm min}}^{9 m_{*,\rm max}} \left(\frac{1}{3}\right)\frac{dn_c}{d\ln m_c} \, dm_c
}
\equiv f_{L,1}(>m_*)
\label{f1l}
\end{equation}
for stellar masses $m_* > m_{\rm frag}/3$. Note the factors of $1/9$ in the numerator and $1/3$ in the denominator, reflecting that, in this scenario, only $1/9$ of the mass in a core of mass $m_c$ is incorporated into a star of mass $m_* = m_c/9$, but that $1/3$ of the core mass goes into stars overall. If we instead adopt the second scenario, where $1/3$ of the core mass goes into $n_{\rm frag}$ stars of mass $m_* = m_c/(3 n_{\rm frag})$, we obtain
\begin{equation}
f(>m_*) = \frac{
\int_{3 n_{\rm frag} m_*}^{3 n_{\rm frag} m_{*,\rm max}} \left(\frac{1}{3}\right)\frac{dn_c}{d\ln m_c} \, dm_c
}{
\int_{3 m_{*,\rm min}}^{3 n_{\rm frag} m_{*,\rm max}} \left(\frac{1}{3}\right)\frac{dn_c}{d\ln m_c} \, dm_c
}
\equiv f_{L,2}(>m_*).
\end{equation}

A final complication is that, in both scenarios, we have implicitly assumed that the core mass function extends to infinity, or at least to $3 n_{\rm frag} m_{*,\rm max} = 1800$ $\msun$. This is not necessarily the case -- the origin of the observed cutoff in the IMF is not understood, and one possible explanation for it is that there simply are no protostellar cores whose mass is larger than a few hundred $\msun$ that are capable of collapsing to single stars even in regions of high surface density. As a simple example of how this would change the results, suppose that there is a maximum core mass $m_{c,\rm max} = 3 m_{*,\rm max} = 360$ $\msun$, sufficient to make a 120 $\msun$ star if the core does not fragment. In this case the mass fractions $f_{L,1}$ and $f_{L,2}$ we have just computed are modified to
\begin{eqnarray}
f_{L,1a}(>m_*) & = & \frac{
\int_{9 m_*}^{m_{c,\rm max}} \left(\frac{1}{9}\right)\frac{dn_c}{d\ln m_c} \, dm_c
}{
\int_{3 m_{*,\rm min}}^{m_{c,\rm max}} \left(\frac{1}{3}\right)\frac{dn_c}{d\ln m_c} \, dm_c
} \\
f_{L,2a}(>m_*) & = & \frac{
\int_{3 n_{\rm frag} m_*}^{m_{c,\rm max}} \left(\frac{1}{3}\right)\frac{dn_c}{d\ln m_c} \, dm_c
}{
\int_{3 m_{*,\rm min}}^{m_{c,\rm max}} \left(\frac{1}{3}\right)\frac{dn_c}{d\ln m_c} \, dm_c
}.
\label{f2la}
\end{eqnarray}

\begin{figure}
\plotone{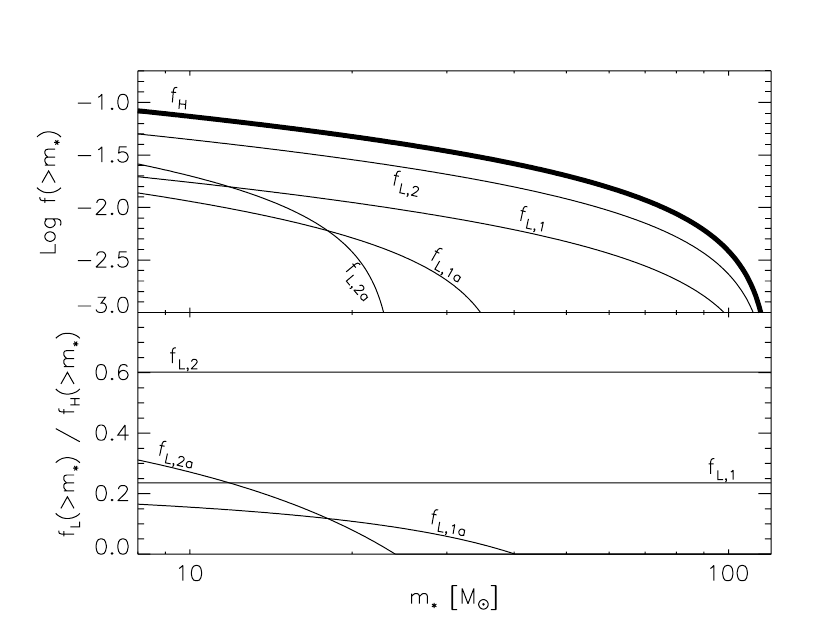}
\caption{
\label{hmfrac}
{\it Top panel:} fraction of mass $f(>m_*)$ in stars with mass greater than $m_*$, for regions of high surface density (thick line, $f_{H}$, Equation \ref{fh}) and regions of low surface density (thin lines, $f_{L,1}$, $f_{L,2}$, $f_{L,1a}$, and $f_{L,2a}$, corresponding to Equations \ref{f1l} -- \ref{f2la}). {\it Bottom panel:} ratio of $f_L(>m_*)/f_H(>m_*)$ massive star mass fraction in low $\Sigma$ regions to that in high $\Sigma$ regions. As in the top panel, $f_L = f_{L,1}$, $f_{L,2}$, $f_{L,1a}$, or $f_{L,2a}$, as indicated.
}
\end{figure}

Evaluating the functions $f_H$, $f_{L,1}$, $f_{L,2}$, $f_{L,1a}$, and $f_{L,2a}$ gives the results shown in Figure \ref{hmfrac}. As the plot shows, reduced star formation efficiency due to the fragmentation in massive cores of the sort we have found can reduce the fraction of total stellar mass in high mass stars by a significant amount. The minimum reduction in massive star fraction, by 40\%, is for $f_{2,L}$, corresponding to the scenario where massive cores fragment into a few equal mass objects and the core mass function extends to infinity. In this case the full mass of cores that fragment is still available to make massive stars, and the massive star fraction declines only because stars of mass $m_*$ must be produced by cores of mass $3 n_{\rm frag} m_*$ in regions of low $\Sigma$ rather than by cores of mass $3 m_*$ in regions of high $\Sigma$, and the total mass of cores available is lower for higher $m_c$.

Any other scenario gives a much more significant reduction in the massive star fraction in regions of low surface density. In scenario $f_{L,1}$, where fragmenting cores produce one massive star of mass $m_c/9$ and only low mass stars otherwise, the reduction is by 76\%. The reduction is partly for the same reason as for $f_{L,2}$, and partly because all the mass in the massive core that does not go into massive stars still goes into low mass stars. Finally, if the core mass function has a cutoff, the reduction in massive star fraction is even more dramatic. In this scenario fragmentation means that a core of mass $m_c = 9 m_*$ (for $f_{L,1a}$) or of mass $m_c = 3 n_{\rm frag} m_*$ (for $f_{L,2a}$) is required to make a star of mass $m_*$, and if this exceeds the maximum core mass then no stars of mass $m_*$ can form in regions of low $\Sigma$.

Regardless of which scenario is ultimately correct, our results show that in regions of low surface density we expect a significant decline in the stellar mass fraction in massive stars compared to a canonical IMF.  The reduction is anywhere from a factor of $1.7$ in the most conservative scenario to a very large factor in more liberal scenarios. This provides numerical confirmation of the hypothesis advanced by \citetalias{krumholz08a} that there is an effective threshold for massive star formation.

\subsection{Implications of Environmental Variation in the IMF}

A variable IMF has numerous implications on scales ranging from the sub-galactic to the cosmological, and a full exploration of them is beyond the scope of this paper. Moreover, a full exploration of this topic would require a larger parameter study than the one we present here, allowing a full exploration of how fragmentation and the stellar mass function vary with environment. Nonetheless, we can identify some important implications.

On the scales of star clusters, preferential formation of massive stars in regions of high surface density suggests that clusters are likely born mass-segregated, with more massive stars forming in the center where the surface density is highest. Outer parts of the cluster, where the surface density drops below $\sim 1$ g cm$^{-2}$, should form preferentially low mass stars. There is some evidence for such primordial mass segregation in Orion \citep{huff06}, but debate continues about whether mass segregation in clusters in general is a result of formation \citep{bonnell98b} or dynamical processes during the first few Myr of cluster lifetime \citep{tan06a, mcmillan07}. Both may occur together, though recent observational \citep{furesz08a, tobin09a} and theoretical \citep{offner08a, offner09b} work pointing out that stars are born sub-virial with respect to their parent clouds (even if the clouds themselves are virialized) suggests that dynamical segregation may be much stronger than earlier estimates found \citep{allison09a}. In very dense clusters, primordial mass segregation may also strongly influence the dynamical evolution of the cluster \citep[e.g.][]{chatterjee09a}. 

Conversely, we do not expect a correlation between the IMF and cluster mass beyond the trivial one expected from finite sampling of a universal IMF, since there does not appear to be a strong correlation between protocluster gas cloud masses and surface densities \citep{fall10a}. A correlation between cluster mass and the IMF has has been suggested by some models \citep[e.g.][]{weidner06}, based on idea that massive stars form via a process in which all gas clouds fragment down to the small initial Jeans mass, but that some subsequently grow through competitive Bondi-Hoyle accretion \citep{bonnell04}. However, the premise on which these models are based -- fragmentation of all clouds down to the Jeans mass at the initial, low cloud temperature -- clearly fails when radiation is included. While the process that goes on in run L might roughly be described as competition, there is clearly no competition in runs M or H, where radiation ensures that most of the proposed competitors never form in the first place. 

Observations remain divided on whether there is a correlation between cluster mass and maximum stellar mass. \citet{weidner04, weidner06} and \citet{weidner09a} claim to detect one, while \citet{oey04a}, \citet{elmegreen06}, and \citet{parker07} argue that there no correlation is present. \citet{dewit04, dewit05} find that $4\pm 2\%$ of galactic O stars formed outside of a cluster of significant mass, which is consistent with the models presented here (for example runs M and H form effectively isolated massive single stars or binaries), but not with the proposed cluster-stellar mass correlation.

No comparable studies have been conducted to search for systematic variation of the IMF with surface density, which we do predict. Such studies are likely to be even more challenging than those searching for a correlation with cluster mass, because cluster surface densities evolve very rapidly once star formation ends and the remaining gas is expelled. For this reason any search for a correlation between IMF and surface density would have to target clusters that are still embedded in their parent gas clouds, for which the cloud surface density can be measured. Unfortunately this renders optical observations, the most common method for determining stellar masses, impossible, since a surface density $\Sigma=1$ g cm$^{-2}$ corresponds to $A_V\approx 200$. Instead other methods of estimating populations of low and high mass stars, such as x-ray observations, would be required \citep{krumholz08a}.

On larger scales, our ability to make predictions is limited by our lack of a theoretical model capable of connecting the surface densities measured over the small scales of star-forming regions where fragmentation occurs ($\sim 1$ pc) to those averaged over much larger areas of galactic disks ($\sim 1$ kpc). Regions of high surface density where massive stars form are clearly found preferentially in regions of high galactic surface density such as spiral arms and near galactic centers, while low surface density clouds such as Taurus are found in regions of lower surface density. However, there is no clear one-to-one mapping from large to small scales. Nonetheless, it seems clear that our results do predict a suppression in the formation of high mass stars in regions where the galactic surface density is low, such as dwarf galaxies and the outer parts of spiral galactic disks. Such a correlation may have been observed \citep{boissier07a, meurer09a, lee09a}, although it too remains controversial \citep{boselli09a}.

\section{Summary}
\label{conclusion}

We report the results of a series of adaptive mesh refinement radiation-hydrodynamic simulations of the collapse of a massive gas cloud using our code ORION. In these simulations, the initial density and velocity structure, the initial virial ratio, and all numerical aspects of the runs are held constant, but the clouds are scaled to different initial surface densities, leading them to respond differently to the radiation feedback produced by the stars that form in the cloud. We find that this differing response leads to dramatic differences in the ways the clouds fragment. A cloud with an initial surface density $\Sigma = 0.1$ g cm$^{-2}$, typical of low-mass star-forming regions such as Taurus or Perseus, fragments strongly, such that it produces a number of stars of comparable mass and puts only a small fraction of its initial mass into the most massive star. No massive stars form in this run. In contrast, runs with initial surface densities of 1 and 10 g cm$^{-2}$, typical of Galactic massive star-forming regions and extra-galactic super-star clusters, respectively, fragment much less. The run with $\Sigma = 1$ g cm$^{-2}$ puts most of its mass into a single massive binary system, while the one with $\Sigma = 10$ g cm$^{-2}$ ends with 90\% of the stellar mass in a single star.

We show that these differing outcomes can be understood in terms of the way that radiation feedback from the stars forming in the cloud affect its subsequent fragmentation. The higher surface density clouds are characterized by higher accretion rates, leading to higher accretion luminosities at early times. Furthermore, their higher optical depths trap the resulting radiation more effectively. The net effect is that radiation feedback raises the temperature and thus the Jeans mass over a significant fraction of the cloud mass in the highest surface density runs, while affecting only much smaller regions when the surface density is low. This leads to an increasing suppression of fragmentation as the initial surface density rises.

Our results suggest that the stellar IMF need not be universal between regions of low surface density ($\Sigma \ll 1$ g cm$^{-2}$) and those of high surface density ($\Sigma \ga 1$ g cm$^{-2}$). In the former, even if turbulence creates the same mass spectrum of initial protostellar cores as in the latter, these cores will fragment during collapse, producing small clusters rather than individual star systems. This effect can dramatically reduce the fraction of the mass of a stellar population in stars with masses $\ga 10$ $\msun$.

\acknowledgements We thank N.~J.\ Evans, C.\ Weidner, and an anonymous referee for helpful comments. Support for this work was provided by: an Alfred P.\ Sloan Fellowship (MRK); NASA through ATFP grant NNX09AK31G (RIK, CFM, and MRK); NASA part of the Spitzer Theoretical Research Program, through a contract issued by the JPL (MRK); the National Science Foundation through grants AST-0807739 (MRK) and AST-0908553 (RIK and CFM); the US Department of Energy at the Lawrence Livermore National Laboratory under contract DE-AC52-07NA 27344 (AC and RIK). Support for computer simulations was provided by an LRAC grant from the National Science Foundation.

\bibliographystyle{apj}
\bibliography{refs}

\end{document}